\documentclass[fleqn,twoside]{article}
\usepackage{espcrc2}

\usepackage{graphicx}

\input epsf
\def\plotone#1{\centering \leavevmode
\epsfxsize= 1.0\columnwidth \epsfbox{#1}}

\def\muK{\mu{\rm K}}
\def\eg {{\it e.g.}}

\def\etal {{\it et al.}}
\def\be{\begin{equation}}
\def\ee{\end{equation}}
\def\bea{\begin{eqnarray}}
\def\eea{\end{eqnarray}}

\def\vecx{{\bf x}}
\def\vecv{{\bf v}}

\def\para{\parallel}
\def\la{\mathrel{\mathpalette\fun <}}
\def\ga{\mathrel{\mathpalette\fun >}}
\def\fun#1#2{\lower3.6pt\vbox{\baselineskip0pt\lineskip.9pt
  \ialign{$\mathsurround=0pt#1\hfil##\hfil$\crcr#2\crcr\sim\crcr}}}

\newcommand{\AmS}{{\protect\the\textfont2
  A\kern-.1667em\lower.5ex\hbox{M}\kern-.125emS}}

\title{Using Galaxy Cluster Peculiar Velocities to Constrain 
Cosmological Parameters}

\author{A. Peel\address[MCSD]{UC Davis Cosmology Group,\\
        1 Shields Ave, Davis, CA 95616, 
           USA}\thanks{peel@physics.ucdavis.edu}
        and
        L. Knox\addressmark\thanks{lknox@ucdavis.edu}}
       
\begin{document}

\begin{abstract} 
Galaxy cluster peculiar velocities can be inferred from high--sensitivity,
high--resolution multiple--frequency observations in the 30 to 400 GHz
range.  While galaxy cluster counts and power spectra are sensitive to the
growth factor, peculiar velocities are sensitive to the time--derivative
of the growth factor and are hence complementary.  Using linear
perturbation theory, we forecast constraints on $\Omega_m$, $H_0$ and the
dark--energy equation of state parameter, $w$, given 820 densely sampled
cluster locations (at $z\simeq 1$) from a $\Lambda$CDM N--body simulation
and 820 sparsely sampled cluster locations in a broader redshift range.
\end{abstract}

\maketitle

\section{Introduction}

After the cosmic microwave background, galaxy clusters may be our best
hope for a precision cosmological probe.  Although clusters are highly
non--linear objects, the dominant importance of gravitational effects over
non--gravitational ones makes them attractive candidates as systems whose
properties should be calculable from first principles.  Cluster number
density evolution has been proposed as a way to determine cosmological
parameters including the dark--energy equation of state parameter
\cite{haiman}.  Here we investigate constraints on cosmology possible from
measurements of cluster peculiar velocities.

Peculiar velocities of galaxies measured via, \eg, the Tully--Fisher and
fundamental plane relations have already been used to constrain
cosmological parameters \cite{jing}.  These efforts have been hindered by
large statistical and systematic errors in the velocity measurements.  In
contrast, cluster peculiar velocities determined from the kinetic and
thermal Sunyaev-Zel'dovich (SZ) effects are potentially more reliable.  
In addition, SZ is the only way to determine peculiar velocities over a
wide range of redshifts allowing one to observe the evolution of
statistical properties.

\section{Sunyaev-Zel'dovich Effects}

About 1\% of the cosmic microwave background (CMB) photons that pass
through the gravitationally confined hot gas in the deep potential wells
of galaxy clusters Compton scatter off a free electron.  On average these
photons gain energy causing a spectral distortion of the blackbody nature
of the CMB characterized by a deficit of low frequency photons and an
excess at high frequencies, with a null near $\nu = 217$ GHz. Specifically
the distortion has the frequency dependence \cite{suny80}:
\bea
\label{eqn:tSZ}
{\Delta I_\nu\over I_\nu} = y \left(x{e^x+1\over e^x-1}-4\right);
\ \ \  x = {h\nu\over kT_{CMB}};\\
y = {k\sigma_T\over m_ec^2}\int dl\ T_e(l)n_e(l) \approx \tau{kT_e\over
m_ec^2}.\nonumber
\eea
This spectral distortion has been observed in the direction of numerous
known clusters, whose locations had been determined by other means
(optical and/or X-ray surveys) \cite{grego}.

The bulk motion of the cluster with respect to the CMB causes the kinetic
SZ effect which, to first order, leads to a hotter (colder)  blackbody
spectrum for a cluster that is moving toward (away) from the observer  
\cite{suny80}:
\be
{\Delta I_\nu\over I_\nu} = -{v_r\over c}{\tau x e^x\over e^x-1} 
\rightarrow
\left({\delta T_{CMB}\over T_{CMB}}\right)_{SZ} = - {v_r\over c}\tau
\ee
where $v_r$ is the radial component of the velocity.

Measurements at multiple frequency bands in the 30 to 400 GHz range can be
used to simultaneously determine $y \simeq \tau {kT_e\over m_ec^2}$ and
$\tau v_r/c$.  These two determinations together with a temperature
determination can be used to solve for $v_r/c$ \cite{suny80}:
\be
{v_r\over c} = -{kT_e\over m_e c^2}{(\delta T_{CMB}/T_{CMB})_{SZ}\over y}.
\ee
For hotter clusters there are temperature--dependent corrections to
Eq.~\ref{eqn:tSZ} that allow one to solve for the temperature as well. For
cooler clusters, X--ray determinations of the temperature may be
necessary.

The best determinations of peculiar velocities from SZ measurements have
errors of $\sigma_v \simeq 1000$ km/s \cite{holzapfel97} but can be
greatly reduced with more sensitive measurements.  Published predictions
of achievable $\sigma_v$ \cite{haehnelt96} assume that the motions of the
electrons in the cluster are simply thermal motions plus one coherent bulk
motion.  In this case there is indeed a well--defined ``velocity of the
cluster'', and the error in its measurement is:
\be
\sigma_v \approx 25\ {\rm {km\over s}} \left(0.01\over\tau\right)
\left({\Delta T_{\rm CMB}^2 + \Delta T_{\rm noise}^2\over 4
\muK^2}\right)^{1\over 2}.
\ee
In the no--noise ($\Delta T_{\rm noise}=0$) limit and with a confusion
noise from the Ostriker--Vishniac effect \cite{ostriker86} of $\Delta
T_{\rm CMB} = 2\ \mu$K (achievable for clusters of small angular extent
which includes any cluster at $z \ga 0.2$, and with high ($\sim 1'$)
angular resolution) then $\sigma_v \simeq 25$ km/sec is possible.

However, clusters have {\it many} bulk flows in them, not just one.
Integrating over all these internal bulk flows, with the mass-weightings
naturally provided by the SZ effects, should still lead to what might be
called ``{\em the} velocity of the cluster''.  But a sizeable fraction of
the mass is along lines of sight with quite low optical depth and hence
high velocity errors.  The net result is that $\sigma_v \simeq 100$ km/sec
is a more likely error \cite{holder}.

\section{Survey Strategy Issues}
 
We imagine peculiar velocities determined as part of a high--resolution,
multi--frequency (30 to 400 GHz) follow--up campaign on a cluster survey.  
These clusters may have been found from X--ray, SZ or even galaxy redshift
surveys (e.g., SDSS and DEEPII).  Important issues are how and how
accurately to determine redshifts, how many velocity measurements to make
at each redshift and whether to sample densely to study correlations or
sparsely to cover more volume.

Cluster redshifts may be obtained via (i) optical photometric and/or
spectroscopic redshifts of galaxy cluster members, (ii) measurement of CO
lines in galaxy cluster members, or (iii) X-ray measurements with
sufficient spectral resolution.  X-ray and optical data also
provide valuable information on the dynamic state of the
cluster; e.g, has the cluster fully virialized, or is it still merging?  
Cluster redshifts are already vitally important for
determining $dN/dz$ from thermal SZ surveys.  Note that interpretation of
peculiar velocity correlations is much less sensitive to selection effects
than the $dN/dz/d\Omega$ statistic.

Dense sampling puts much higher demand on the accuracy of the redshift
determinations.  For a specified error in comoving separation $\Delta r$
we need a redshift error smaller than $\Delta z =0.003 h (H(z)/H_0)  
(\Delta r/(10\ {\rm Mpc}))$; in order to calculate the expected
correlations accurately we need $\Delta r \la 10$ Mpc.  With sparse
sampling, all we need is a coarse redshift binning and assurance that the
clusters are separated by $\ga 200$ Mpc (comoving); photometric redshifts
are almost certainly sufficient for that case.

There are good arguments for targeting the $z \simeq 0.2$ to 0.5 range
rather than higher redshifts.  The tolerance on redshift error decreases
only slowly with increasing redshift, whereas the amount of telescope time
required typically increases dramatically.  In addition, optical and
X--ray data on the clusters may be valuable for investigating the
dynamical state of the cluster and third X--ray temperatures are necessary
for the colder clusters.

\section{Linear Theory}

The evolution of the Fourier--transformed density contrast $\delta_k$ in
the linear regime is separable, allowing us to write:
\be
\delta_k(\eta) = D(\eta)\delta_k(\eta_0)
\ee
where $D(\eta)$ is the growth factor and $\eta_0$ is the conformal time
today.  From the continuity equation for matter, $ikv_k = \dot{\delta}_k$,
we see that velocities are probes of the time derivative of $D$.

\begin{figure}[htb]
\plotone{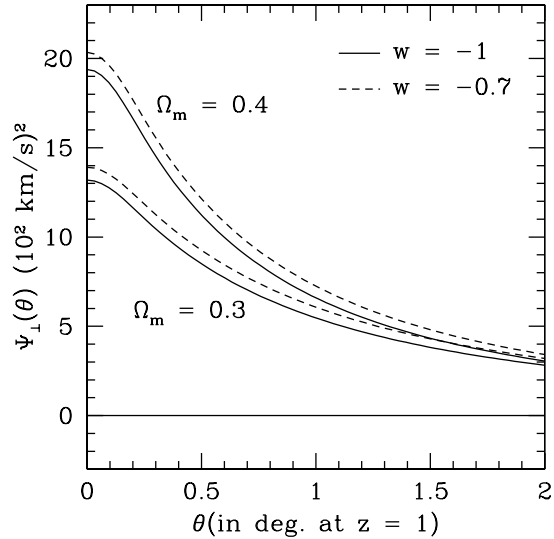}
\caption{Linear peculiar velocity correlations for pairs of clusters at 
redshift $z = 1$ as a function of separation angle.
\label{fig:psiperp}}
\end{figure}

The correlation between radial velocity components of two clusters at
locations $(r_i,\hat{ \gamma}_i)$ and $(r_j,\hat{ \gamma}_j)$ relative to
the observer is given by \cite{gorski}:
\bea
\label{eqn:psi}
\Psi_{ij} & \equiv & \langle\hat{\gamma}_i\cdot\vecv(\vecx_i)
\vecv(\vecx_j)\cdot\hat{\gamma}_j\rangle \\
& = & \Psi_\perp\cos\theta + (\Psi_\para - 
\Psi_\perp)R(\theta,r_i,r_j)\nonumber \\
R(\theta,r_i,r_j)
& \equiv & {(r_i^2 + r_j^2)\cos\theta - r_ir_j(1 + \cos^2\theta)
\over r_i^2 + r_j^2 - 2r_ir_j\cos\theta} \nonumber\\
\Psi_{\perp,\para} & = & {\dot{D}(r_i)\dot{D}(r_j)\over 2\pi^2} \int dk\
|\delta_k|^2\ K_{\perp,\para}(kr)\nonumber
\eea
where $\cos\theta = \hat{\gamma}_i\cdot\hat{\gamma}_j$, $K_\perp (kr) =
j_1(kr)/(kr)$, $K_\para (kr) = j_0(kr) - 2j_1(kr)/(kr)$, $r$ is the
comoving distance between the clusters and an overdot symbolizes
$d/d\eta$. Figure \ref{fig:psiperp} shows $\Psi_\perp(\theta)$ for four
different flat cosmologies at redshift $z=1$.

Cosmology dependence in Eq.~\ref{eqn:psi} arises from: (i) the
redshift--distance relation $r(z)$ and (ii) the time--derivative of the
growth factor, $\dot D$.  The latter has weak dependence on $w$ as we can
see from the highly accurate analytic approximation \cite{wangstein}:
\be
\dot{D}(z)  = D(z)a(z)H(z) [\Omega_m(z)]^{\alpha(w)}
\ee
because $\alpha$ only ranges from about 6/11 to 3/5 for $w=-1$ to $w=0$, 
and because $a(z)D(z)H(z)$ is also fairly insensitive to $w$ for $z \la 
0.5$.  In contrast, $\dot{D}$ is highly sensitive to $\Omega_m$.

\section{Forecasted Parameter Errors}

For nearly uncorrelated cluster velocities, from, \eg, 
a very sparse survey, we can easily estimate the expected error variance
on $\Omega_m$ from measurement of N clusters with expected 
peculiar velocity variance $\Psi_0(z_i)$:
\bea
(\Delta \Omega_m)^2 =  \left(\sum_i \left({\partial \Psi_0(z_i) \over 
\partial \Omega_m}
\right)^2 {1 \over 2(\Psi_0(z_i)+\sigma_v^2)^2 }\right)^{-1}\nonumber
\eea
\be
\ \ \ \ \ \ \ \ \ \ \ \simeq  {800\over N}(.01)^2
\ee
where the last equality assumes $N$ clusters with $\sigma_v^2\ll \Psi_0$ 
all at $z=1$, and $\partial \ln \Psi_0 /\partial \Omega_m \simeq 5$. 

We now turn to a Fisher matrix analysis of two survey types. For a dense
survey, we sample 820 neighboring clusters from the Hubble Volume
Lightcone Simulation Cluster Catalog \cite{virgo} over an area of 80 sq.
deg. at redshift $z\simeq 1$.  For a sparse survey, we construct a grid of
820 clusters from $z=0.1$ to 1.0 over an area of 2600 sq. deg such that
all comoving cluster separations are greater than 200 Mpc.

\begin{figure}[htb]
\plotone{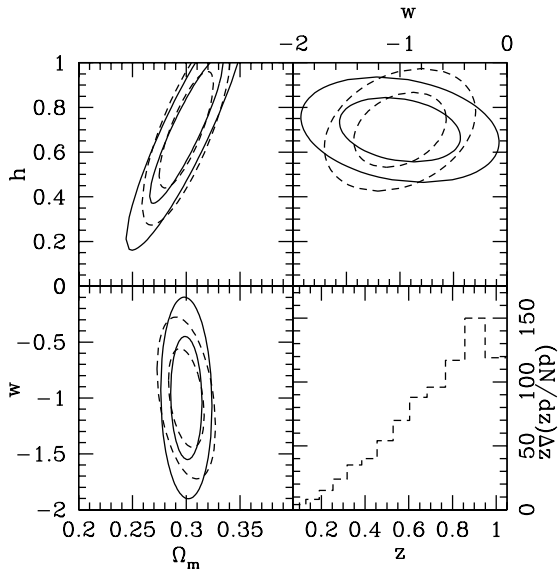}
\caption{Forecasted 1 and 2--$\sigma$ contours for the dense survey
(solid) and the sparse survey with $(dN/dz)\Delta z$ as shown (dashed).  
For each two--parameter contour, the third parameter is held fixed.  [The
drop in $dN/dz$ in the final bin is due to a quirk in our sampling
algorithm.] 
\label{fig:enchilada}}
\end{figure}

We calculate the expected covariances in linear theory for a density field
smoothed with a 5 Mpc radius tophat filter.  We calculate their partial
derivatives by finite difference.  The models we difference have fixed
COBE normalization \cite{bunn}.  All our results in Figure
\ref{fig:enchilada} assume a noise level of 100 km/s.

As expected, $\Omega_m$ is measured very well.  Neither $h$ nor $w$ are
well constrained; nor do they lead to significant confusion in the
$\Omega_m$ measurement.  In addition the results are very similar for our
two different survey types, with the sparse survey (dashed lines)
marginally more constraining than the dense survey.  Both increasing the
velocity errors to greater than the assumed 100 km/sec and including the
effects of redshift errors will improve the sparse sampling result in
comparison to the dense sampling result.

\section{Non--linear and biasing effects}

We expect that linear theory provides only a rough guide to the
constraints possible from peculiar velocity measurements.  
Numerically--determined variances agree with linear theory predictions to
$\sim$ 40\% in the present epoch \cite{colberg00}.  We plan to include
non--linear and biasing effects by use of the PINOCCHIO gravitational
instability code \cite{monaco}.  With PINOCCHIO we will obtain highly
accurate calculations of the velocity correlation functions in several
models and once again use finite difference to calculate Fisher matrices.

\section{Conclusions}

Cluster peculiar velocity measurements are highly sensitive to $\Omega_m$
and largely insensitive to $h$ and $w$. Sparse sampling is probably
preferable to dense sampling. Velocity determinations of $\sim 800$
clusters can potentially be used to determine $\Omega_m$ to $\sim 0.03\%$.  
This $w$--independent measurement of $\Omega_m$ may be useful for
improving $w$ constraints from observations of type Ia supernovae
\cite{perl}.

We would like to thank B. Holden, G. Holder, M. Kaplinghat, C.-P. Ma, J. 
Mohr, S. Meyer, M. White, G. Wilson and I. Zehavi for very useful 
conversations.


\begin{thebibliography}{99}

\bibitem{haiman} Z. Haiman, J. Mohr, G. Holder (2001) ApJ {\bf 553}: 545

\bibitem{jing} \eg, Y.P. Jing, G. B\"orner, Y. Suto (2002) ApJ {\bf 564}:  
15; S. Zaroubi, \etal, (2001) MNRAS {\bf 326}: 375

\bibitem{suny80} R.A. Sunyaev, Ya.B. Zel'dovich (1980) Sov. Astron. Lett.
{\bf 6}: 737;
M. Birkinshaw (1998) Phys. Reprts. {\bf 310}: 97

\bibitem{grego} L. Grego \etal (2001) ApJ {\bf 552}: 2

\bibitem{holzapfel97} W. Holzapfel, \etal, (1997) ApJ {\bf 479}: 17; W.
Holzapfel, et al. (1997) ApJ {\bf 481}: 35

\bibitem{haehnelt96} N. Aghanim, K. G\'orski, J.-L. Puget (2001) A\&A {\bf
374}: 1-12; M.G. Haehnelt, M. Tegmark (1996) MNRAS {\bf 279}: 545

\bibitem{ostriker86} J.P. Ostriker, E.T. Vishniac (2000) ApJL {\bf 306}:
51; W. Hu, M. White (1997) ApJ {\bf 479}: 568; A.H. Jaffe, M.  
Kamionkowski (1998) Phys. Rev. D {\bf 58}: 043001; V. Springel, M. White, 
L. Hernquist (2001) ApJ {\bf 549}: 681; C.-P. Ma, J. Fry (2002)  PRL
{\bf 88}: 211301

\bibitem{holder} G. Holder (2002) in preparation.

\bibitem{gorski} K. G\'orski (1988) ApJ {\bf 332}: L7

\bibitem{wangstein} L. Wang, P. Steinhardt (1998) ApJ {\bf 508}: 483

\bibitem{virgo} J.M. Colberg, \etal, (The Virgo Consortium) (2000), 
MNRAS {\bf 319}: 209

\bibitem{bunn} E. Bunn, M. White (1997) ApJ {\bf 480}: 6

\bibitem{colberg00} J.M. Colberg, \etal, (The Virgo Consortium) (2000)
MNRAS {\bf 313}: 229

\bibitem{monaco} P. Monaco, T. Theuns, G. Taffoni (2002) MNRAS {\bf 
331}: 587

\bibitem{perl} S. Perlmutter \etal, (1999) ApJ {\bf 517}: 565

\end{thebibliography}
\end{document}